# Political Events using RAG with LLMs

Muhammad Arslan[a]*, Saba Munawar[b] and Christophe Cruz[a]

[a]Laboratoire Interdisciplinaire Carnot de Bourgogne, Université de Bourgogne, Dijon, France
[b]National University of Computer and Emerging Sciences (NUCES), Islamabad, Pakistan

**Abstract**

In the contemporary digital landscape, media content stands as the foundation for political news analysis, offering invaluable insights sourced from various channels like news articles, social media updates, speeches, and reports. Natural Language Processing (NLP) has revolutionized Political Information Extraction (IE), automating tasks such as Event Extraction (EE) from these diverse media outlets. While traditional NLP methods often necessitate specialized expertise to build rule-based systems or train machine learning models with domain-specific datasets, the emergence of Large Language Models (LLMs) driven by Generative Artificial Intelligence (GenAI) presents a promising alternative. These models offer accessibility, alleviating challenges associated with model construction from scratch and reducing the dependency on extensive datasets during the training phase, thus facilitating rapid implementation. However, challenges persist in handling domain-specific tasks, leading to the development of the Retrieval-Augmented Generation (RAG) framework. RAG enhances LLMs by integrating external data retrieval, enriching their contextual understanding, and expanding their knowledge base beyond pre-existing training data. To illustrate RAG's efficacy, we introduce the Political EE system, specifically tailored to extract political event information from news articles. Understanding these political insights is essential for remaining informed about the latest political advancements, whether on a national or global scale.

## 1. Introduction

In today's digital era, media content stands as a valuable resource for political scientists, providing crucial insights into political landscapes, events, and sentiments [1]. This content, spanning news articles, social media posts, speeches, and reports, provides insight into public opinion, political trends, and policy implications [2].

---

* Corresponding author. Tel.: +33 03 80 39 50 00.
E-mail address: muhammad.arslan@u-bourgogne.fr

However, manual methods for extracting political information from media are increasingly impractical due to the sheer volume of digital data [3]. The rise of NLP has transformed IE from media, enabling automated analysis of text data to classify, categorize, and interpret large collections of textual information [4]. NLP techniques have significantly enhanced the efficiency and accuracy of IE tasks in political science, covering event detection, hate speech identification, sentiment analysis, and more [5]. Traditional NLP methods require specialized expertise and extensive resources to build rule-based systems or train machine learning models with domain-specific datasets. However, the emergence of LLMs driven by GenAI (i.e., encompasses AI tools designed for creating diverse content types like text, images, and audio that closely resembles human-produced material) offers accessible, open-source solutions, reducing the challenges of building models from scratch and minimizing the need for extensive datasets [6, 7]. Despite LLMs improving, they still struggle with domain-specific tasks, so RAG was developed [8]. RAG aims to make LLMs more practical by pulling in external data during the generative process, boosting their real-world effectiveness.

RAG with LLMs is widely used in many areas, but it is not significantly explored in politics yet [9, 10]. There is a substantial lack of research on how it could improve the IE process and enhance political information systems. To fill this void, we present the integration of RAG with the LLM to create a Political EE system. The system offers several benefits: improved accuracy and efficiency in event extraction, immediate updates for users to stay informed, and enhanced credibility through attribution of information sources. By combining generative capabilities with external knowledge retrieval, the system provides comprehensive and reliable insights into the political landscape, facilitating informed decision-making and analysis for researchers and policymakers.

The paper is structured as follows: Section 2 delves into the relevant literature on Political EE and explores current applications of RAG with LLMs. Section 3 provides an overview of the system, demonstrating the integration of RAG with LLMs for political EE. In Section 4, the evaluation of the system is presented. Section 5 discusses the strengths and weaknesses of the system. Finally, Section 6 offers concluding remarks on the paper.

## 2. Background

Table 1 summarizes various use cases of EE and analysis in political contexts [11 – 26], utilizing diverse datasets. These use cases include analyzing protest events over time and space, identifying triggers for state-led mass killings, and detecting political events through syntax and semantics. Specific scenarios cover event detection in smart cities, linking events and locations in political texts, and geocoding battle events. The datasets used range from news articles, Twitter data, and Wikipedia to LexisNexis, English Gigaword documents, ACE 2005, and the UCLA protest image dataset. These applications leverage a variety of techniques and models. Traditional approaches include keyword search, location-based filters, and document classifiers. Advanced machine learning techniques such as Support Vector Machines (SVMs), Recurrent Neural Networks (RNNs), Convolutional Neural Networks (CNNs), and Latent Dirichlet Allocation (LDA) are also employed [27 – 29]. Classification methods like Multinomial Naive Bayes, Complement Naive Bayes, and Random Forest are identified as well. Neural-net based models, the Stanford CoreNLP toolkit, and Cosine similarity are essential tools in this domain. Despite the benefits of these NLP techniques, choosing and validating the appropriate systems for specific needs can be challenging due to limited access to data and resources. Smaller organizations, in particular, may struggle with the time and investment required for Research and Development (R&D). To address this, utilizing pre-trained and readily accessible LLMs is recommended.

LLMs represent a significant milestone in the field of NLP, showcasing unprecedented capabilities in understanding and generating human-like text across diverse domains [30]. These models, trained on massive datasets, possess a remarkable capacity for contextual understanding and language generation, making them invaluable tools for a wide range of NLP tasks. However, while LLMs excel in general applications, they often encounter challenges when tasked with domain-specific inquiries or when confronted with limited data availability. RAG emerges as a transformative solution to address these challenges by seamlessly integrating LLMs with external data retrieval mechanisms [8]. By augmenting LLMs with access to diverse sources of information during the generative process, RAG enhances their contextual understanding, relevance, and accuracy [8]. Numerous studies have explored the versatility of RAG across various domains. Han et al. [31] analyze tutoring practices using RAG with middle-school dialogue transcripts to aid educational decision-making. Alawwad et al. [32] examine RAG's effectiveness in handling out-of-domain scenarios, particularly in textbook QA across life, earth, and physical science lessons. Bucur et al. [33] employ RAG for automated form filling to enhance enterprise search functionalities. Zhang et al. [34] investigate RAG for financial sentiment analysis using Twitter news and FiQA datasets. Al Ghadban et al. [35] utilize RAG for frontline health worker capacity building, focusing on pregnancy-related guidelines for health education QA. Jeong et al. [36] introduce Self-BioRAG, a framework for biomedical text analysis. Xia et al. [37] propose a hybrid RAG model for real-time composition assistance, incorporating diverse datasets to enhance writing speed and accuracy. Rackauckas [38] employs RAG-Fusion for technical product information QA, while Shi et al. [39] introduce RACE, a system for commit message generation in code intelligence using multiple programming languages.

After reviewing the mentioned studies, it is evident that combining RAG with LLMs shows significant promise in extracting structured data across diverse fields. This integration enhances contextual comprehension and ensures data accuracy, providing valuable insights for decision-making. However, despite its success in various domains, its application in politics remains relatively unexplored. Specifically, the potential of this fusion to refine the EE process and enhance political information systems warrants further investigation. To address this gap, we propose exploring the integration of RAG with LLMs to develop a Political EE system.

Table 1. Existing research on political event extraction and analysis.

| No. | Use case | Details | Dataset used |
| --- | --- | --- | --- |
| 1 | Protest event analysis over time and space [11] | Events such as strikes, demonstrations, petitions, etc. | News articles |
| 2 | Systematic analysis of triggers of state-led mass killings [12] | Trigger-type events: escalations of armed conflict, armed conflict spillover from a neighboring country, etc. | News articles |
| 3 | Political events using syntax and semantics [13] | Events such as torture, prison, killing, harassment, discrimination, etc. | News articles |
| 4 | Event detection in smart cities [14] | Events such as thefts, car accidents, traffic jams, electricity charges, flood risks, etc. | Twitter data |
| 5 | Linking events and locations in political text [15] | Reported offensives in Syria (e.g., civilian deaths) | News articles and Wikipedia |
| 6 | Acquisition of patterns for coding political event data [16] | Protest events | LexisNexis data service, English Gigaword documents, ACE 2005, etc. |
| 7 | Coding political events [17] | Producing multilingual event data. | Wikipedia data and Gigaword text corpus |
| 8 | Detecting battle events [18] | Geocoding armed conflicts events on the Russo-Ukrainian war. | Telegram data |
| 9 | Event prediction by few-shot abductive reasoning [19] | Integrating a large language model in event prediction. | Political events, interactions between social political entities and user reviews. |
| 10 | Event detection from real-time streaming data [20] | Classifying events into local or global events. | Twitter data |
| 11 | Multilingual protest events detection [21] | Analyzing multilingual, cross-lingual, few-shot, and zero-shot settings for socio-political event information collection. | News articles |
| 12 | Protest event identification [22] | Multilingual events | News articles |
| 13 | Multilingual socio-political and crisis event detection [23] | Extracting document and cross-sentence level events. | News articles |
| 14 | Civil unrest event detection [24] | Jointly identifies relevant tweets and detects civil unrest events. | Twitter data |
| 15 | Protest event detection [25] | Recognizing protesters and estimating the level of perceived violence. | UCLA protest image dataset |
| 16 | Pattern matching for event detection [26] | Disease outbreaks and terrorism-related events. | News snippets |

## 3. Political EE system

This section explores the details of political events, explaining their definition and presenting the various attributes that contribute to their contextual understanding. According to Davenport and Ball [40], political events encompass actions undertaken by political actors within specific temporal and spatial contexts. To facilitate this understanding, the proof-of-concept demonstration adopts the event property schema outlined by Halterman [41], which describes eight potential properties (see Fig. 1). These properties include the "actor" identifying the agent performing the action; the "action" describing the action itself; the "recipient" providing information about the actor receiving the action; and the "instrument," encompassing objects or means used to carry out the action. Additionally, the "reason" property explains the cause of the reported event, while the "time" and "location" properties denote when and where the event occurred, respectively. Finally, the "reporter" property identifies the source that reported the event. While an event necessitates at least one action and one actor or recipient, the inclusion of other properties is optional, and sentences containing all eight pieces of information are relatively rare [41].

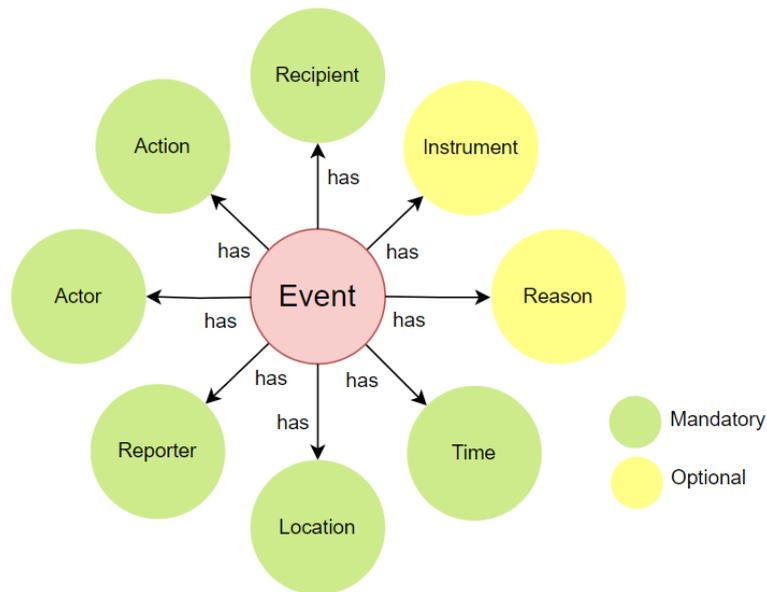

Fig. 1. Properties of political events.

Having identified the key properties essential for constructing political events, we integrate the power of the RAG with LLM model, driven by GenAI, to automatically extract these events without the need for manual intervention in extracting each event property. To facilitate interactive exploration of political events, we opt to develop a Political EE system. Leveraging the expertise of LLMs in complex reasoning tasks, we explore several publicly available LLM releases, including GPT-4 [42], Llama2 [43], etc. For our proof-of-concept demonstration, we opted for LLama2 from the range of available LLMs. However, other LLMs are also viable options for exploration. To tailor the selected LLM specifically for political tasks and to effectively harness political EE from news articles, we have integrated RAG technology into our solution. This strategic fusion significantly enhances the system's capability to understand and address political-related queries using the extracted information. The system's ability to handle a variety of political queries depends on the quality and comprehensiveness of the provided news dataset. We chose to work with the News Category Dataset [44], which comprises news headlines published between 2012 and 2022, filtered to include only data from 2020 to 2022 for system development purposes. While this dataset is old, it offers comprehensive coverage of news articles related to political events. To develop the system (see Fig. 2), we utilized Algorithm 1, establishing a seamless connection between LLama2 and the dataset, thereby facilitating the creation of an RAG application.

First, news articles and questions are provided as input (see Algorithm 1). The process begins by installing necessary packages like Transformers [45], langchain (langchain.com), and llama_index (llamaindex.ai), followed by importing required modules such as VectorStoreIndex and HuggingFaceLLM. System prompts and query wrapper prompts are defined for LLama2 to provide context and structure for the language model. Next, the algorithm logs in to Hugging Face using huggingface-cli to access the models and initializes the HuggingFaceLLM model. It then initializes LangchainEmbedding and creates a ServiceContext, choosing an embedding model to convert text into vectors and managing embeddings, LLama2, and other components. Documents are loaded from a specified directory using SimpleDirectoryReader, and embeddings are generated to build a VectorStoreIndex. The query_engine is then used to search the index for relevant information based on the provided question, leveraging the LLama2 model and embeddings to retrieve accurate answers. Finally, the response generated by the query_engine is obtained and further processed or displayed as needed, producing text that answers the posed question.

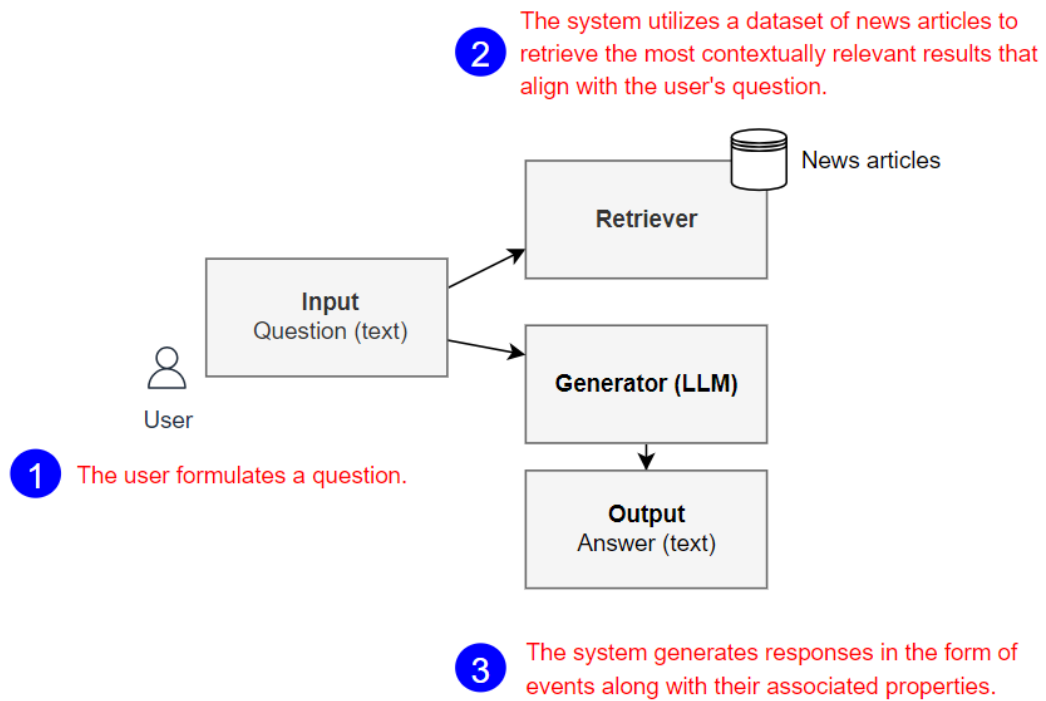

Fig. 2. The RAG with LLM-based system is utilized by a user who employs an external dataset of news articles to generate a response.

## 4. System Evaluation

The primary objective of evaluating the political EE system, constructed on the foundation of RAG with LLM model, was to enhance efficiency in extracting political events. The evaluation primarily focused on assessing the system's capability to accurately identify named entities, including individuals and organizations, while also extracting relevant properties associated with political events. To create a comprehensive test dataset, we manually curated a selection of 50 health-related political events from the dataset, with a representative subset outlined in Table 2. Subsequently, we conducted entity-to-entity comparisons utilizing the Cosine Similarity matrix, a method that computes the cosine of the angle between two vectors to gauge similarity in a high-dimensional space. Throughout the evaluation phase, the system consistently demonstrated robust performance, achieving an impressive accuracy rate of 0.87. This notable outcome underscores the effectiveness of the RAG-LLM model in proficiently extracting political events and accurately identifying named entities within health-related political contexts.

Algorithm 1: Implementing RAG with LLM for political EE.

| |
|---|
| **Objective: Political events analysis using LLama2 and Hugging Face Models.** |
| **1) Input**: News articles, question. |
| **2) Process**: |
|    **2.1 Install necessary packages:** |
|       Transformers, langchain, sentence_transformers, llama_index, llama-index-llms-huggingface, llama-index-embeddings-langchain |
|    **2.2 Import required modules and packages:** |
|       VectorStoreIndex, HuggingFaceLLM, SimpleInputPrompt, VectorStoreIndex, SimpleDirectoryReader |
|    **2.3 Define system_prompt and query_wrapper_prompt for LLama2. These prompts provide context and structure for the language model.** |
|    **2.4 Log in to Hugging Face using huggingface-cli to access models from the Hugging Face model hub.** |
|    **2.5 Initialize HuggingFaceLLM.** |
|    **2.6 Initialize LangchainEmbedding and create ServiceContext:** |
|       - Choose an embedding model (e.g., HuggingFaceEmbeddings) to convert text into vectors. |
|       - Create a ServiceContext to manage embeddings, LLama2, and other components. |
|    **2.7 Create VectorStoreIndex from documents:** |
|       - Load documents from a specified directory using SimpleDirectoryReader. |
|       - Utilize the ServiceContext to generate embeddings for each document and build the index. |
|    **2.8 Query the index with a specific question:** |
|       - Use the query_engine to search for relevant information based on the provided question. |
|       - The query engine leverages the LLama2 model and embeddings to retrieve accurate answers. |
|    **2.9 Obtain the response generated by the query_engine, which contains relevant information extracted from the documents.** |
|    **2.10 Further process or display the response as needed for the application or user interface.** |
| **3) Output:** The text generated as the answer to the question posed. |

Table 1. Sampe records of user queries and system generated answers for system evaluation.

| Year | Health-Related Events with Dates and Authors |
|---|---|
| 2022 | 1. Senate Republicans Yield, Aid Passage of Veterans Health Bill Following Prior Obstruction (Date: 2022-08-02, Author: Jennifer Bendery) |
| | 2. GOP Shifts Stance on Deficit, Prioritizing Fertility Over Fiscal Responsibility (Date: 2022-07-15, Author: Jonathan Nicholson) |
| | 3. Supreme Court Faces Protests Over Leaked Draft on Abortion Rights (Date: 2022-05-03, Author: Marita Vlachou) |
| | 4. FDA Greenlights First COVID-19 Breathalyzer Test (Date: 2022-04-14, Author: Sarah Ruiz-Grossman) |
| | 5. Biden's COVID Coordinators Stepping Down, Ashish Jha to Take Over (Date: 2022-03-17, Author: Unnamed) |
| | 6. Idaho Governor Candidate Ammon Bundy Arrested in Baby Seizure Protest (Date: 2022-03-13, Author: Mary Papenfuss) |
| | 7. Queen Elizabeth Contracts COVID-19 (Date: 2022-02-20, Author: Unnamed) |
| 2021 | 8. Judge overturns Texas Bans on School Mask mandates in blow to GOP Governor (Date: 2021-11-11, Author: Nick Visser) |
| | 9. Washington Governor Mandates Vaccines for All School Employees. (Date: 2021-08-19, Author: Nick Visser) |
| | 10. Trump's Muslim Ban Harmed Muslim Americans' Health, Study Finds (Date: 2021-08-04, Author: Rowaida Abdelaziz) |
| | 11. Florida Again Leads Nation in Soaring COVID Cases Amid Delta Fears (Date: 2021-07-25, Author: Mary Papenfuss) |
| | 12. Democrats Introduce Bill to Invest in Public Safety Alternatives to Police (Date: 2021-06-28, Author: Sarah Ruiz-Grossman) |
| | 13. Judge Blocks CDC from Enforcing Cruise Ship Coronavirus Rules (Date: 2021-06-19, Author: Sara Boboltz) |
| | 14. More than 1 Million Have Signed Up for Coverage at Healthcare.gov This Year (Date: 2021-05-11, Author: Jonathan Cohn) |
| | 15. Fauci: 'There's No Doubt' COVID-19 Deaths have been Undercounted in U.S. (Date: 2021-05-09, Author: Nina Golgowski) |

## 5. Discussion

The Political EE system, designed to extract political events through the fusion of RAG with LLM (specifically, the Llama2 model in our implementation), presents several key advantages. Firstly, it streamlines development by avoiding the trial-and-error process typically associated with constructing political EE tasks using classical NLP techniques. Leveraging pre-trained LLMs like Llama2 facilitates the rapid development of tailored political IE solutions with minimal cost and resource overhead, particularly advantageous for organizations with limited R&D budgets. Secondly, the system enables seamless dynamic updates to the domain dataset, allowing for effortless adaptation to the latest information by substituting outdated data files with up-to-date ones. This ensures constant tracking of new political events without interruption. Additionally, the system enhances the credibility and reliability of extracted information by attributing information sources through RAG, instilling confidence in users regarding the authenticity of the provided insights.

However, the study encounters several limitations. While LLama2 was chosen, exploring different LLMs tailored for the political domain could yield valuable insights. The objective of this research was not to select the best LLM for political EE, but rather to showcase the integration of RAG with LLMs for understanding political events. Additionally, the system utilized an outdated dataset of news articles to demonstrate its functionality, chosen primarily due to its online availability. However, using the latest news dataset would better reflect the system's performance. Furthermore, the system operated on a small dataset to showcase its functionality, but employing extensive datasets with RAG may lead to slower response times for event extraction. This issue could potentially be mitigated by scaling up computational resources, such as increasing the number of Graphics Processing Units (GPUs), although at a considerable cost for companies with limited resources. Further research is warranted in this domain. Moreover, while the built system focused on utilizing news articles as the primary data source, political information originates from various channels such as social media (e.g., Twitter), government websites, etc. RAG should seamlessly integrate these diverse information sources to extract comprehensive and precise political insights. To achieve this, multi-source RAG architectures need to be explored. As news from online sources is a continuous process, it may not be optimal to scrape and integrate information manually into the system. Hence, real-time RAG solutions should be investigated to enable analysts to observe political dynamics in real-time.

## 6. Conclusion

IE plays a pivotal role in political science, transitioning from traditional methodologies to advanced machine learning techniques driven by third-generation approaches, particularly utilizing NLP for context-aware document analysis. LLMs have significantly expedited this progression, facilitating swift and accurate EE from news articles. Building upon this foundation, our study introduces the Political EE system, designed to assist analysts in navigating the complexities of political IE. By integrating RAG with LLM, the system transcends the limitations of classical NLP techniques, providing a streamlined approach to development devoid of the trial-and-error process. Notably, it offers dynamic dataset updates, enabling continuous adaptation to the latest political events, thereby ensuring uninterrupted exploration. Moreover, the system enhances the credibility and reliability of extracted information by attributing sources through RAG, adding confidence in users regarding the authenticity of the insights provided. As we move forward, the fusion of RAG with LLM holds immense potential to further revolutionize political EE, offering a robust framework for comprehensive and precise analysis in the dynamic landscape of political science.

## Acknowledgements

The authors express gratitude to the French Government for funding provided by the National Research Agency (ANR).

## References

[1] Papacharissi Z. (2015) "Affective publics: Sentiment, technology, and politics." Oxford University Press.
[2] Gollust SE, Fowler EF, Niederdeppe J. (2019) "Television news coverage of public health issues and implications for public health policy and practice." *Annual Review of Public Health* **1**(**40**):167-85.

[3] Giuffrida G, Gozzo S, Rinaldi FM, Tomaselli V. (2017) "Extracting Info From Political News Through Big Data Network Analysis." *Rivista Italiana di Economia Demografia e Statistica* **71**(**1**).

[4] Piskorski J, Yangarber R. (2013) "Information extraction: Past, present and future. Multi-source, multilingual information extraction and summarization.":23-49.

[5] Bamman D, Smith NA. (2015) "Open extraction of fine-grained political statements." InProceedings of the 2015 conference on empirical methods in natural language processing: 76-85.

[6] Huang K, Huang G, Dawson A, Wu D. (2024) "GenAI Application Level Security." *Generative AI Security: Theories and Practices*: 199-237. Cham: Springer Nature Switzerland.

[7] Yao Y, Duan J, Xu K, Cai Y, Sun Z, Zhang Y. (2024) "A survey on large language model (llm) security and privacy: The good, the bad, and the ugly." High-Confidence Computing:100211.

[8] Lewis P, Perez E, Piktus A, Petroni F, Karpukhin V, Goyal N, Küttler H, Lewis M, Yih WT, Rocktäschel T, Riedel S. (2020) "Retrieval augmented generation for knowledge-intensive nlp tasks." *Advances in Neural Information Processing System* **33**:9459-74.

[9] Huang Y, Huang J. (2024) "A Survey on Retrieval-Augmented Text Generation for Large Language Models." arXiv preprint arXiv:2404.10981.

[10] Haq EU, Braud T, Kwon YD, Hui P. (2020) "A survey on computational politics." IEEE Access 8:197379-406.

[11] Lorenzini J, Kriesi H, Makarov P, Wüest B. (2022) "Protest event analysis: Developing a semiautomated NLP approach." American Behavioral Scientist **66**(**5**):555-77.

[12] Burley T, Humble L, Sleeper C, Sticha A, Chesler A, Regan P, Verdeja E, Brenner P. (2020) "Nlp workflows for computational social science: Understanding triggers of state-led mass killings." *Practice and Experience in Advanced Research Computing*: 152-159.

[13] Halterman, A. (2020). "Extracting political events from text using syntax and semantics." Technical report MIT.

[14] Hodorog A, Petri I, Rezgui Y. (2022) "Machine learning and Natural Language Processing of social media data for event detection in smart cities." *Sustainable Cities and Society* **1**(**85**):104026.

[15] Halterman, A. (2018). "Linking Events and Locations in Political Text." MIT Political Science Department Research Paper No. 2018-21, Available at SSRN: https://ssrn.com/abstract=3267476 or http://dx.doi.org/10.2139/ssrn.3267476

[16] Makarov, P. (2018, August). Automated acquisition of patterns for coding political event data: two case studies. In Proceedings of the second joint SIGHUM workshop on computational linguistics for cultural heritage, social sciences, humanities and literature (pp. 103-112).

[17] Liang, Y., Jabr, K., Grant, C., Irvine, J., & Halterman, A. (2018, July). New techniques for coding political events across languages. In 2018 IEEE International Conference on Information Reuse and Integration (IRI) (pp. 88-93). IEEE.

[18] Tanev, H., Stefanovitch, N., Halterman, A., Uca, O., Zavarella, V., Hürriyetoğlu, A., ... & Della Rocca, L. (2023, September). Detecting and geocoding battle events from social media messages on the russo-ukrainian war: Shared task 2, case 2023. In Proceedings of the 6th Workshop on Challenges and Applications of Automated Extraction of Socio-political Events from Text (pp. 160-166).

[19] Shi, X., Xue, S., Wang, K., Zhou, F., Zhang, J., Zhou, J., ... & Mei, H. (2024). Language models can improve event prediction by few-shot abductive reasoning. Advances in Neural Information Processing Systems, 36.

[20] Singh, J., Pandey, D., & Singh, A. K. (2023). Event detection from real-time twitter streaming data using community detection algorithm. Multimedia Tools and Applications, 1-28.

[21] Hürriyetoğlu, A., Mutlu, O., Yörük, E., Liza, F. F., Kumar, R., & Ratan, S. (2021). Multilingual protest news detection-shared task 1, case 2021. In Proceedings of the 4th Workshop on Challenges and Applications of Automated Extraction of Socio-political Events from Text (CASE 2021) (pp. 79-91).

[22] Suri, M., Chopra, K., & Arora, A. (2022). NSUT-NLP at CASE 2022 Task 1: Multilingual Protest Event Detection using Transformer-based Models. In Proceedings of the 5th Workshop on Challenges and Applications of Automated Extraction of Socio-political Events from Text (CASE) (pp. 161-168).

[23] Hettiarachchi, H., Adedoyin-Olowe, M., Bhogal, J., & Gaber, M. M. (2021). DAAI at CASE 2021 task 1: Transformer-based multilingual socio-political and crisis event detection. In Proceedings of the 4th Workshop on Challenges and Applications of Automated Extraction of Socio-political Events from Text (CASE 2021) (pp. 120-130).

[24] Delucia, A., Dredze, M., & Buczak, A. L. (2023). A multi-instance learning approach to civil unrest event detection on twitter. In Proceedings of the 6th Workshop on Challenges and Applications of Automated Extraction of Socio-political Events from Text (pp. 18-33).

[25] Won, D., Steinert-Threlkeld, Z. C., & Joo, J. (2017). Protest activity detection and perceived violence estimation from social media images. In Proceedings of the 25th ACM international conference on Multimedia (pp. 786-794).

[26] Tanev, H. (2024). Leveraging approximate pattern matching with bert for event detection. In Proceedings of the 7th Workshop on Challenges and Applications of Automated Extraction of Socio-political Events from Text (CASE 2024) (pp. 32-39).

[27] Steinwart, I., & Christmann, A. (2008). Support vector machines. Springer Science & Business Media.

[28] Blei, D. M., Ng, A. Y., & Jordan, M. I. (2003). Latent dirichlet allocation. Journal of machine Learning research, 3(Jan), 993-1022.

[29] Grossberg, S. (2013). Recurrent neural networks. Scholarpedia, 8(2), 1888.

*Arslan et al. / Procedia Computer Science 00 (2024) 000–000*  9
[30] Ge Y, Hua W, Mei K, Tan J, Xu S, Li Z, Zhang Y. (2024) "Openagi: When llm meets domain experts." *Advances in Neural Information Processing Systems* **13**;36.
[31] Han, Z. FeiFei, Lin, J., Gurung, A., Thomas, D. R., Chen, E., Borchers, C., Gupta, S., & Koedinger, K. R. (2024). "Improving Assessment of Tutoring Practices using Retrieval-Augmented Generation." arXiv preprint arXiv:2402.14594.
[32] Alawwad, H. A., Alhothali, A., Naseem, U., Alkhathlan, A., & Jamal, A. (2024). "Enhancing Textbook Question Answering Task with Large Language Models and Retrieval Augmented Generation." arXiv preprint arXiv:2402.05128.
[33] Bucur, M. (2023). "Exploring Large Language Models and Retrieval Augmented Generation for Automated Form Filling" (Bachelor's thesis, University of Twente).
[34] Zhang B, Yang H, Zhou T, Ali Babar M, Liu XY. (2023) "Enhancing financial sentiment analysis via retrieval augmented large language models." InProceedings of the Fourth ACM International Conference on AI in Finance: 349-356.
[35] Al Ghadban, Y., Lu, H. Y., Adavi, U., Sharma, A., Gara, S., Das, N., ... & Hirst, J. E. (2023). "Transforming healthcare education: Harnessing large language models for frontline health worker capacity building using retrieval-augmented generation." *medRxiv*, 2023-12.
[36] Jeong, M., Sohn, J., Sung, M., & Kang, J. (2024). "Improving Medical Reasoning through Retrieval and Self-Reflection with Retrieval-Augmented Large Language Models." arXiv preprint arXiv:2401.15269.
[37] Xia, M., Zhang, X., Couturier, C., Zheng, G., Rajmohan, S., & Ruhle, V. (2023). "Hybrid retrieval-augmented generation for real-time composition assistance." arXiv preprint arXiv:2308.04215.
[38] Rackauckas, Z. (2024). "RAG-Fusion: a New Take on Retrieval-Augmented Generation." arXiv preprint arXiv:2402.03367.
[39] Shi, E., Wang, Y., Tao, W., Du, L., Zhang, H., Han, S., ... & Sun, H. (2022). "RACE: Retrieval-Augmented Commit Message Generation." arXiv preprint arXiv:2203.02700.
[40] Davenport C, Ball P. (2002) "Views to a kill: Exploring the implications of source selection in the case of Guatemalan state terror, 1977-1995." *Journal of conflict resolution* **46**(**3**):427-50.
[41] Halterman, A. (2021). "Three Essays on Natural Language Processing and Information Extraction with Applications to Political Violence and International Security" (Doctoral dissertation, Massachusetts Institute of Technology).
[42] Achiam, J., Adler, S., Agarwal, S., Ahmad, L., Akkaya, I., Aleman, F. L., ... & McGrew, B. (2023). "Gpt-4 technical report." arXiv preprint arXiv:2303.08774.
[43] Touvron, H., Martin, L., Stone, K., Albert, P., Almahairi, A., Babaei, Y., ... & Scialom, T. (2023). "Llama 2: Open foundation and fine-tuned chat models." arXiv preprint arXiv:2307.09288.
[44] Misra, R. (2022). "News category dataset." arXiv preprint arXiv:2209.11429.
[45] Wolf, T., Debut, L., Sanh, V., Chaumond, J., Delangue, C., Moi, A., ... & Rush, A. M. (2020, October). Transformers: State-of-the-art natural language processing. In Proceedings of the 2020 conference on empirical methods in natural language processing: system demonstrations (pp. 38-45).